\documentclass[journal=jpcbfk,manuscript=article]{achemso}

\usepackage[version=3]{mhchem} 
\usepackage{tikz}
\usepackage{amsmath,amssymb}

\def\xa{{x_{\alpha}}}
\def\xb{{x_{\beta}}}
\def\ab{{\alpha\beta}}
\def\gab{{g_{\alpha\beta}}}
\def\rmIdeal{{\rm ideal}}
\def\rmInfo{{\rm info}}
\def\rmFluct{{\rm fluct}}
\def\rmElec{{\rm elec}}
\def\rmAve{{\rm ave}}
\def\rmChem{{\rm chem}}
\def\rmd{{\rm d}}
\def\br{{\bf r}}
\def\bp{{\bf p}}
\def\kB{{k_{\rm B}}}

\author{M. C. Gao}
\affiliation[National Energy Technology Laboratory and AECOM]
{National Energy Technology Laboratory, Albany OR  97321, USA; AECOM, P.O. Box 1959, Albany OR 97321, USA}
\author{M. Widom}
\email{widom@cmu.edu}
\phone{412-268-7645}
\affiliation[Carnegie Mellon University]
{Department of Physics, Carnegie Mellon University, Pittsburgh, PA  15217, USA}

\title[Information entropy]
{Information Entropy of Liquid Metals}

\begin{document}
\begin{abstract} 
Correlations reduce the configurational entropies of liquids below their ideal gas limits.  By means of first principles molecular dynamics simulations, we obtain accurate pair correlation functions of liquid metals, then subtract the mutual information content of these correlations from the ideal gas entropies to predict the absolute entropies over a broad range of temperatures. We apply this method to liquid aluminum and copper and demonstrate good agreement with experimental measurements, then we apply it to predict the entropy of a liquid aluminum-copper alloy. Corrections due to electronic entropy and many-body correlations are discussed.
\end{abstract}

\section{Introduction}

The remarkable equivalence of information and entropy, as recognized by Shannon~\cite{Shannon1948} and Jaynes~\cite{Jaynes1957}, implies that the atomic coordinates of a solid or liquid contain all the information that is needed to calculate its configurational entropy.  Qualitatively, ordered structures are fully described with little information. For example, specifying a crystal lattice and its atomic basis uniquely determines the positions of infinitely many atoms in a crystallographic structure using a finite amount of information, so the entropy per atom vanishes. Meanwhile a disordered structure requires separately specifying information about each atom, which implies a finite entropy per atom. For example, to specify the distribution of chemical species in a random equiatomic binary solid solution requires $\log_2 2 = 1$ bit of information for each atom. The principle also holds for gases and liquids, with suitable modification to account for continuous positional degrees of freedom as outlined below. Thus, configurational entropies depend on configurations alone and do not require separate knowledge of the interatomic interactions, in contrast to energies which require both the configurations {\em and} the interactions.

Given a distribution of discrete states $i$ with probabilities $p_i$, the expected information required to specify the actual state is~\cite{Shannon1948}
\begin{equation}
\label{eq:Shannon}
S/\kB=-\sum_i p_i\ln{p_i}.
\end{equation}
By choosing the natural logarithm and assigning units of $\kB$ we identify the information as entropy, as suggested by von Neumann~\cite{Petz2001}.  In quantum statistical mechanics~\cite{Neumann1927} we take the Boltzmann probability distribution, $p_i=\exp{(-E_i/\kB T)}/Z$, with the partition function $Z$ as the normalizing factor. Classically, the distribution becomes continuous. In the canonical ensemble the $N$-particle entropy in volume $V$ becomes~\cite{Green1952}
\begin{equation}
\label{eq:S_class}
S_N/\kB=-\frac{1}{N!} \int \prod_i \rmd\br_i\rmd\bp_i~ f_N\ln{(h^{3N}f_N)}.
\end{equation}
where $f_N(\br_1,\bp_1,\dots,\br_N,\bp_N)$ is the $N$-body probability density, as a function of the atomic positions $\br_i$ and momenta $\bp_i$. This expression, including the factors of Planck's constant $h$, can be derived as the high temperature limit of the quantum expression Eq.~(\ref{eq:Shannon}).

Applying Eq.~(\ref{eq:S_class}) to an uncorrelated fluid of density $\rho=N/V$ yields the entropy per atom of the classical ideal gas~\cite{McQuarrie},
\begin{equation}
\label{eq:S_ideal}
S_\rmIdeal/\kB= \frac{5}{2}-\ln{(\rho\Lambda^3)}.
\end{equation}
The $5/2$ term in Eq.~(\ref{eq:S_ideal}) includes $3/2$ coming from three-dimensional integrals of the single-body Maxwell-Boltzmann momentum distribution $f_1(\bp)=\rho(2\pi m \kB T)^{-3/2}\exp{(-\bp^2/2m\kB T)}$, plus an additional $2/2=1$ arising from the second term in the Stirling approximation $\ln{N!}\approx N\ln{N}-N$.  The quantum de~Broglie wavelength $\Lambda=\sqrt{h^2/2\pi m \kB T}$ diverges at low $T$, so this classical $S_\rmIdeal$ approaches $-\infty$. However, the quantization of energy levels in a finite volume yields the low temperature limit $S\rightarrow 0$ as $T\rightarrow 0$~\cite{Kirkwood1933,Kirkwood1934,McQuarrie}. Thus $S_\rmIdeal$ is an {\rm absolute} entropy, consistent with the conventional choice of S=0 at T=0.\footnote{Still, $S_\rmIdeal\rightarrow -\infty$ when we take the thermodynamic limit of infinite volume prior to the low temperature limit $T\rightarrow 0$. However, the ideal gas is not a suitable model for real matter at low temperature. More realistic models with a low density of states at low energy ({\em e.g.} harmonic solids) exhibit vanishing low temperature entropy, consistent with the usual third law of thermodynamics~\cite{Griffiths1965,WidomSM}.}

Equation~(\ref{eq:S_class}) can be reexpressed in terms of $n$-body distribution functions~\cite{Green1952,Yvon1969,Evans1989}, $g_N^{(n)}$, as
\begin{equation}
\label{eq:s1s2s3}
S_N/N\kB=s_1+s_2+s_3+\dots,
\end{equation}
with the $n$-body terms
\begin{align}
\label{eq:s1_canonical}
s_1 &= \frac{3}{2}-\ln{(\rho\Lambda^3)}, \\
\label{eq:s2_canonical}
s_2 &= -\frac{1}{2}\rho\int\rmd\br~ g_N^{(2)}\ln{g_N^{(2)}}, \\
\label{eq:s3_canonical}
s_3 &= -\frac{1}{6}\rho^2\int\rmd\br^2 g_N^{(3)}
                       \ln{(g_N^{(3)}/g_N^{(2)}g_N^{(2)}g_N^{(2)})}.
\end{align}
The subscripts $N$ indicate that the correlation functions are defined in the canonical ensemble with fixed number of atoms $N$. Equations~(\ref{eq:s1_canonical}-\ref{eq:s3_canonical}) appear superficially similar to a virial-type low density expansion. However, we use correlation functions that are nominally exact, not their low density virial approximations, so in fact the series is an expansion in cumulants of the many-body probability distribution.  Truncation of the series is accurate if a higher many-body correlation function can be approximated by the products of fewer-body correlations. For example, the Kirkwood superposition approximation $g^{(3)}(1,2,3)\approx g^{(2)}(1,2) g^{(2)}(2,3) g^{(2)}(1,3)$ causes $s_3$ to vanish.

Mutual information measures how similar a joint probability distribution is to the product of its marginal distributions~\cite{Cover2006}. In the case of a liquid structure, we may compare the two-body joint probability density~\cite{McQuarrie,Rowlinson1982} $\rho^{(2)}(\br_1,\br_2)=\rho^2 g(|\br_2-\br_1|)$ with its single-body marginal, $\rho(\br)$. The mutual information
\begin{equation}
\label{eq:MI}
I[\rho^{(2)}(\br_1,\br_2)]=\frac{1}{N}\int\rmd\br_1\rmd\br_2~
\rho^{(2)}(\br_1,\br_2)\ln{(\rho^{(2)}(\br_1,\br_2)/\rho(\br_1)\rho(\br_2))}
\end{equation}
tells us how much information $g(r)$ gives us concerning the positions of atoms at a distance $r$ from another atom.  Mutual information is nonnegative definite.  We recognize the term $s_2$ in Eq.~(\ref{eq:s2_canonical}) as the negative of the mutual information, with the factor of $1/2$ correcting for double-counting of pairs of atoms.  Thus $s_2$ reduces the liquid state entropy relative to $s_1$ by the mutual information content of the radial distribution function $g(r)$.

Pair correlation functions for liquid metals obtained through {\em ab-initio} molecular dynamics (AIMD) simulation can predict the configurational entropy through Eqs.~(\ref{eq:s1_canonical}-\ref{eq:s3_canonical}), truncated at the two-body level. We demonstrate this method for liquid aluminum and copper, showing good agreement with experimentally measured absolute entropies over broad ranges of temperature. Corrections to the entropy due to electronic excitations and three-body correlations are discussed. Finally, we apply the method to a liquid aluminum-copper alloy.

\section{Theoretical methods}
\label{sec:methods}

\subsection{Entropy expansion}
\label{subsec:entropy}

Direct application of the formalism Eqs.~(\ref{eq:s1_canonical}-\ref{eq:s3_canonical}) is inhibited by constraints such as
\begin{equation}
\label{eq:constraint}
\rho^n\int\prod_i\rmd\br_i~ g_N^{(n)}=\frac{N!}{(N-n)!}
\end{equation}
that lead to long-range (large $r$) contributions to the two- and three-body integrals.  Nettleton and Green~\cite{Nettleton1958}, and Raveche~\cite{Raveche1971a,Raveche1971b}, recast the distribution function expansion in the {\em grand} canonical ensemble and obtained expressions that are better convergent. We follow Baranyai and Evans~\cite{Evans1989} and utilize the constraint~(\ref{eq:constraint}) to rewrite the two-body term as
\begin{equation}
\label{eq:s2}
s_2 = \frac{1}{2}+\frac{1}{2}\rho\int\rmd\br~ [g^{(2)}-1]
      -\frac{1}{2}\rho\int\rmd\br~ g^{(2)}\ln{g^{(2)}}.
\end{equation}
The combined integrand $\{[g^{(2)}(r)-1]-g^{(2)}(r)\ln{g^{(2)}(r)}\}$ falls off rapidly, so that the sum of the two integrals converges rapidly as the range of integration extends to large $r$. Furthermore, the combined integral is ensemble invariant, which allows us to substitute the grand canonical ensemble radial distribution function $g(r)$ in place of the canonical $g_N^{(2)}$. The same trick applies to the three-body term,
\begin{equation}
\label{eq:s3}
s_3 = \frac{1}{6}+\frac{1}{6}\rho^2\int\rmd\br^2 
[g^{(3)}-3g^{(2)}g^{(2)}+3g^{(2)}-1]
-\frac{1}{6}\rho^2\int\rmd\br^2 g^{(3)}\ln{(g^{(3)}/g^{(2)}g^{(2)}g^{(2)})}.
\end{equation}

In the grand canonical ensemble, the first two terms in Eq.~(\ref{eq:s2}) arise from fluctuations in the number of atoms, $N$, and can be evaluated in terms of the isothermal compressibility $\kappa_T$. We define 
\begin{equation}
\label{eq:S_fluct}
\Delta S_\rmFluct[g(r)]/\kB \equiv \frac{1}{2}+\frac{1}{2}\rho\int\rmd\br~[g(r)-1]
                        = \frac{1}{2}\rho \kB T\kappa_T,
\end{equation}
and note that it is positive definite.  The remaining term is the entropy reduction due to the two-body correlation. As noted above, the mutual information content of the radial distribution function $g(r)$ reduces the entropy by
\begin{equation}
\label{eq:I}
\Delta S_\rmInfo[g(r)]/\kB \equiv -\frac{1}{2}\rho\int\rmd\br~ g(r)\ln{g(r)}.
\end{equation}
The complete two-body term is now $s_2=\Delta S_\rmFluct/\kB+\Delta S_\rmInfo/\kB$. The corresponding three-body term in Eq.~(\ref{eq:s3}) reduces to a difference of three- and two-body entropies, and its sign is not determined.

Notice the constant term $5/2$ in the ideal gas entropy, $S_\rmIdeal$ (Eq.~(\ref{eq:S_ideal})), while the one-body entropy, $s_1$ (Eq.~(\ref{eq:s1_canonical})), instead contains $3/2$. The contribution of $1/2$ in $s_2$ as given by Eq.~(\ref{eq:s2}), together with an added $1/6+\cdots=1/2$ from the three-body and higher terms, reconciles the one-body entropy with the ideal gas. For consistency with previous workers~\cite{Nettleton1958,Raveche1971a,Raveche1971b,Evans1989}, and to make connection with the ideal gas, we could add the entire series $1/2+1/6+\cdots = 1$ to $s_1$ and write
\begin{equation}
\label{eq:S_grand}
S_N/N\kB=S_\rmIdeal/\kB + (s_2-1/2)+(s_3-1/6)+\cdots
\end{equation}
which is equivalent to Eq.~(\ref{eq:s1s2s3}).

\subsection{Ab-initio molecular dynamics simulation}

To provide the liquid state correlation functions needed for our study we perform {\em ab-initio} molecular dynamics (AIMD) simulations based on energies and forces calculated from first principles electronic density functional theory (DFT). AIMD provides a suitable compromise between accuracy and computational efficiency. It accurately predicts liquid state densities and pair correlation functions with no adjustable parameters or empirical interatomic interactions.  We apply the plane-wave code VASP~\cite{Kresse99} in the PBEsol generalized gradient approximation~\cite{Perdew08}, utilizing a single $k$-point in a simulation cell of 200 atoms.

Simulations are performed at fixed volume for each temperature. In the case of Cu we fix the volumes at the experimental values~\cite{CRC2011}. Because experimental values are not available over the needed temperature range for Al, and are not available at all for AlCu, we determined these volumes by the condition that the average total pressure (including the kinetic term) vanishes. The predicted volumes for Al are insensitive to the energy cutoff of our plane-wave basis set, so we use the default value of 240 eV. Over the temperature range where volumes are available for Al~\cite{CRC2011}, we reproduce the experimental values to within 0.5\%.  For AlCu an elevated energy cutoff was required. We found 342 eV, which is 25\% above the default for Cu of 273 eV, to be sufficient to achieve convergence.  Given a suitable volume, the energy cutoff has minimal impact on our simulated correlation functions and predicted entropies.

Pair correlation functions are collected as histograms in 0.01~\AA~bins and subsequently smeared with a Gaussian of width $\sigma$=0.025~\AA. Our run durations for data collection were 50~ps for Al and Cu, and 20~ps for AlCu. All structures were thoroughly equilibrated prior to data collection.

\subsection{Electronic entropy}

The electronic density of states $D(E)$, which comes as a byproduct of first principles calculation, determines the electronic entropy. At low temperatures, all states below the Fermi energy $E_F$ are filled and all states above are empty. At finite temperature, single electron excitations vacate states below $E_F$ and occupy states above, resulting in the Fermi-Dirac occupation function
\begin{equation}
f_T(E)=\frac{1}{\exp{[(E-\mu)/\kB T]}+1}
\end{equation}
($\mu$ is the electron chemical potential). Fractional occupation probability creates an electronic contribution to the entropy,
\begin{equation}
\label{eq:S_elec}
\Delta S_\rmElec/\kB = -\int D(E)[f_T(E)\ln{f_T(E)}+(1-f_T(E))\ln{(1-f_T(E))}].
\end{equation}
We apply this equation to representative configurations drawn from our liquid metal simulations, with increased $k$-point density (a $2\times 2\times 2$ Monkhorst grid) in order to converge the density of states.

At low temperatures, the electronic entropy approaches $(\pi^2/3)D(E_F)\kB^2T$, which depends only on the density of states at the Fermi level. However, at the high temperatures of liquid metals, the electronic entropy requires the full integral as given in Eq.~(\ref{eq:S_elec}), rather than its low temperature approximation.

\section{Results and discussion}

\subsection{Application to pure liquid metals}

\begin{figure}
\includegraphics[width=5in,angle=-90]{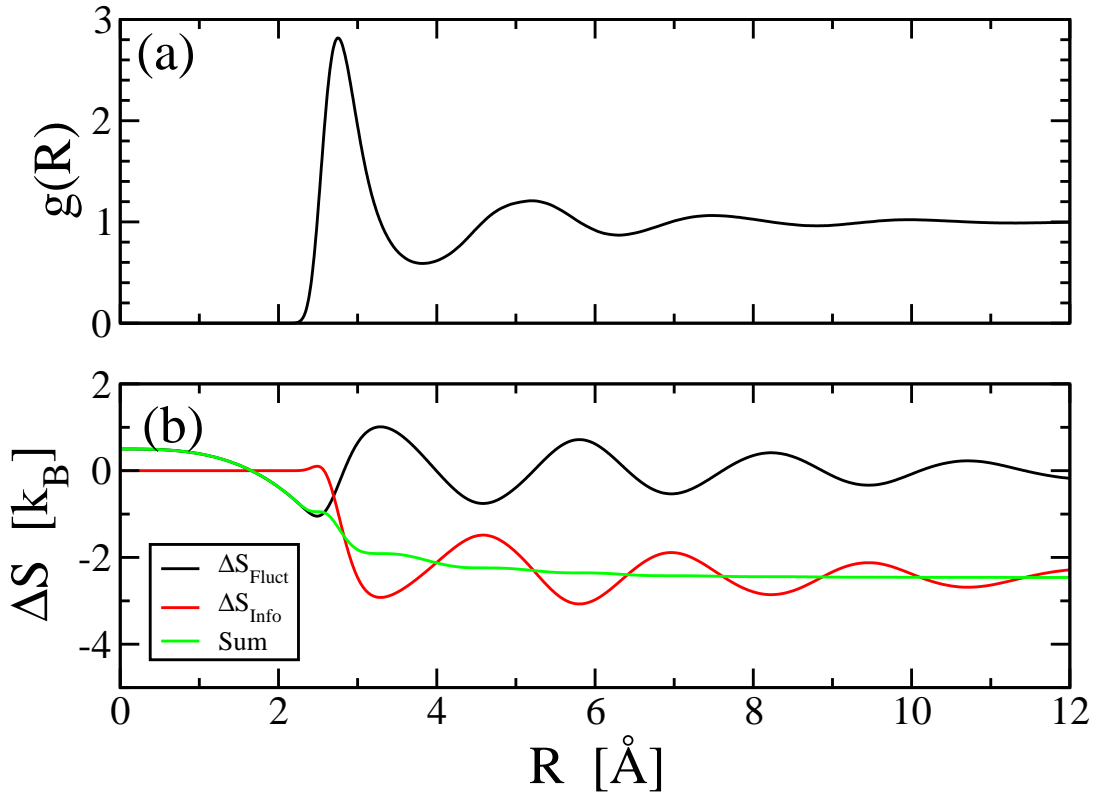}
  \caption{(a) Radial distribution function $g(r)$ of liquid Al at T=1000 K. (b) Contributions to the entropy of liquid Al integrated from $r=0$ up to $R$.}
\label{fig:dS_Al}
\end{figure}

Figure~\ref{fig:dS_Al}a displays a simulated radial distribution function $g(r)$ for liquid Al at T=1000K.  Integrated contributions to the entropy are shown in Fig.~\ref{fig:dS_Al}b. The excluded volume region below 2~\AA, where $g(r)$ vanishes, does not contribute to $\Delta S_\rmInfo$, but it does contribute, negatively, to $\Delta S_\rmFluct$.  Strong peaks with $g(r)>1$ contribute positively to $\Delta S_\rmFluct$. They also contribute positively to mutual information, and hence negatively to $\Delta S_\rmInfo$, reducing the entropy. Minima with $g(r)<1$ do the opposite. The information and fluctuation integrals each oscillate strongly and converge slowly, while their sum is monotone and rapidly convergent.  Note the asymptotic value of $\Delta S_\rmFluct$ is close to zero, as is expected for a liquid metal with low compressibility (for liquid Al, values of $\rho \kB T\kappa_T\sim 0.03-0.04$ are reported~\cite{Jakse2013}). In contrast, the entropy loss due to mutual information is more than $2\kB$.

\begin{figure}
\includegraphics[width=5in,angle=-90]{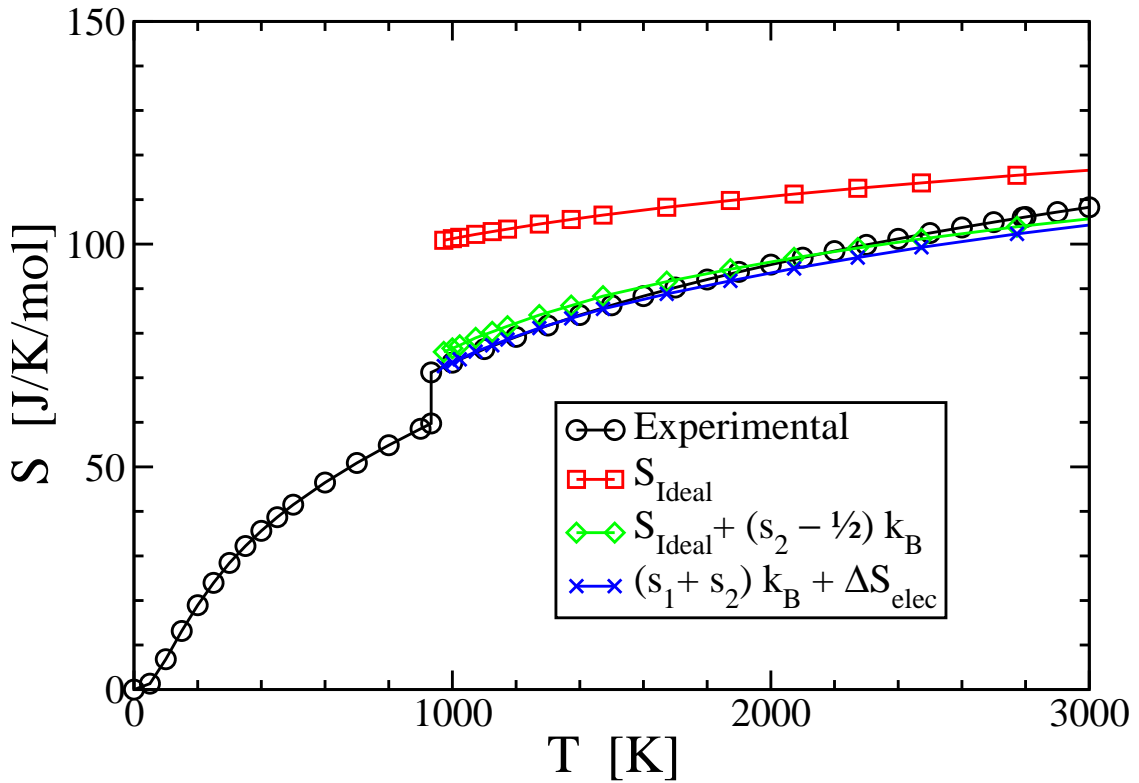}
  \caption{Entropy of liquid Al, comparing the experimental values with various approximations: the ideal gas (Eq.~(\ref{eq:S_ideal})); ideal gas with pair corrections (Eq.~(\ref{eq:S_grand})); single-body entropy with pair correction and electronic entropy, $(s_1+s_2)\kB+\Delta S_\rmElec$.}
\label{fig:S_Al}
\end{figure}

Repeating this calculation at several temperatures, and choosing the values of $\Delta S_\rmFluct$ and $\Delta S_\rmInfo$ obtained at $R=12$~\AA, we predict the absolute entropy as a function of temperature as displayed in Fig.~\ref{fig:S_Al}. Our predictions lie close to experimental values~\cite{Hultgren,Janaf} over the entire simulated temperature range, however there are systematic discrepancies. Our value is too high at low temperatures, and too low at high temperatures. Including a further correction due to electronic entropy (not shown) improves the agreement at high temperature while worsening it at low. As noted in the discussion surrounding Eq.~(\ref{eq:S_grand}), we have arbitrarily included the constants $1/2, 1/6, \dots$ belonging to the fluctuation terms such as Eq.~(\ref{eq:S_fluct}), in the ideal gas entropy. Removing those terms, and instead plotting $(s_1+s_2)\kB+\Delta S_\rmElec$, we find excellent agreement at low T and slight underestimation at high T.

\begin{figure}
\includegraphics[width=5in,angle=-90]{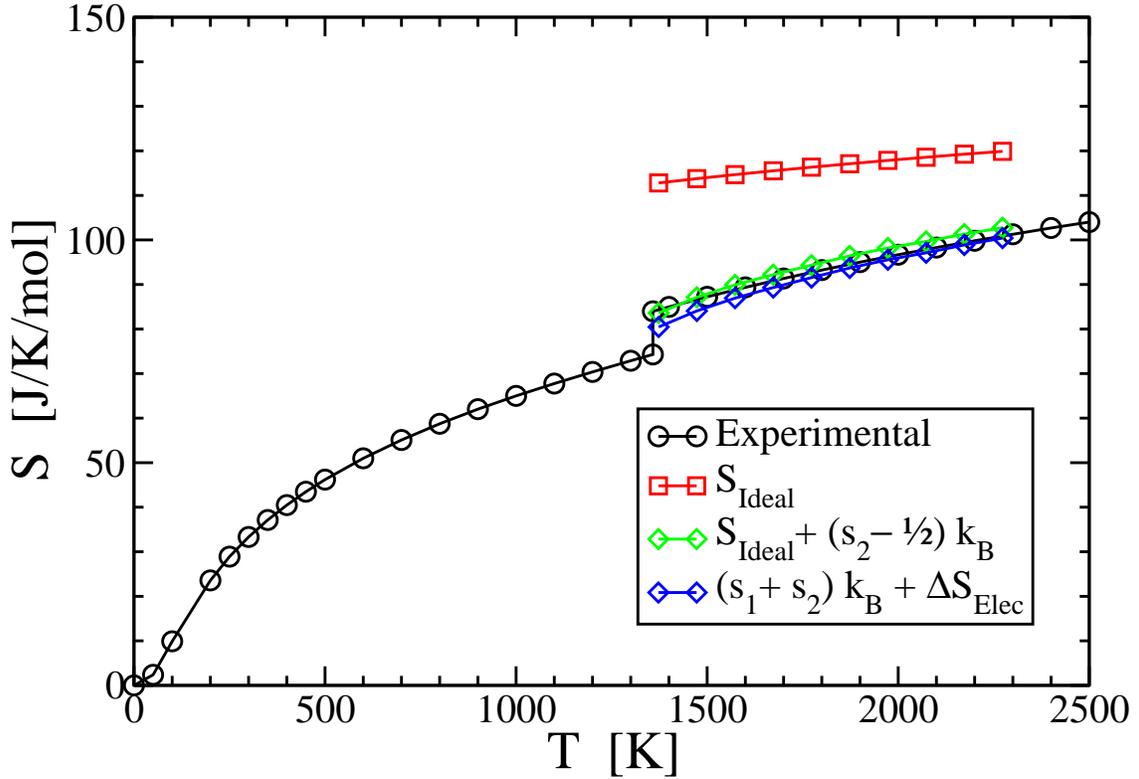}
  \caption{Entropy of liquid Cu, comparing the experimental values with various approximations: the ideal gas (Eq.~(\ref{eq:S_ideal})); ideal gas with pair corrections (Eq.~(\ref{eq:S_grand})); single-body entropy with pair correction and electronic entropy, $(s_1+s_2)\kB+\Delta S_\rmElec$.}
\label{fig:S_Cu}
\end{figure}

We find rather similar behavior in the case of liquid copper (see Fig.~\ref{fig:S_Cu}).  Here, the agreement with experiment is less close, especially at low temperatures.  Presumably we must include many-body corrections such as $s_3$ or higher that are likely to be stronger at low temperatures. The $d$-orbitals of copper lie close to the Fermi surface, possibly causing deviations from the Kirkwood superposition approximation that increase the value of $s_3$.  Excitations of $d$-electrons also contribute significantly to $\Delta S_\rmElec$ causing a faster than linear increase with T.

\subsection{Application to binary AlCu liquid alloy}

Finally, we turn to a liquid aluminum-copper alloy.  As demonstrated by Hernando~\cite{Hernando1990} and applied by Laird and Haymet~\cite{Laird1992b}, Eqs.~(\ref{eq:S_fluct}) and~(\ref{eq:I}) generalize naturally to multicomponent systems, with mole fraction $\xa$ for species $\alpha$ and with partial pair distribution functions $\gab(r)$ between species $\alpha$ and $\beta$:
\begin{equation}
\label{eq:S_fluct_multi}
\Delta S_\rmFluct[\gab(r)]/\kB = \frac{1}{2}+
\frac{1}{2}\rho\sum_\ab\xa\xb\int\rmd\br~[\gab(r)-1],
\end{equation}
and
\begin{equation}
\label{eq:S_info_multi}
\Delta S_\rmInfo[\gab(r)]/\kB = 
-\frac{1}{2}\rho\sum_\ab\xa\xb\int\rmd\br~ \gab(r)\ln{\gab(r)}.
\end{equation}
We set $s_2=\Delta S_\rmFluct/\kB+\Delta S_\rmInfo/\kB$ as before. We also need to revise the ideal gas entropy:
\begin{equation}
\label{eq:S_ideal_multi}
S_\rmIdeal/\kB=\frac{5}{2}-\sum_\alpha\xa\ln{\left(\rho\xa\Lambda_\alpha^3\right)}.
\end{equation}
Notice that the ideal mixing entropy $-\kB\sum_\alpha \xa\ln{\xa}$ is included in this expression for $S_\rmIdeal$.

\begin{figure}
\includegraphics[width=5in,angle=-90]{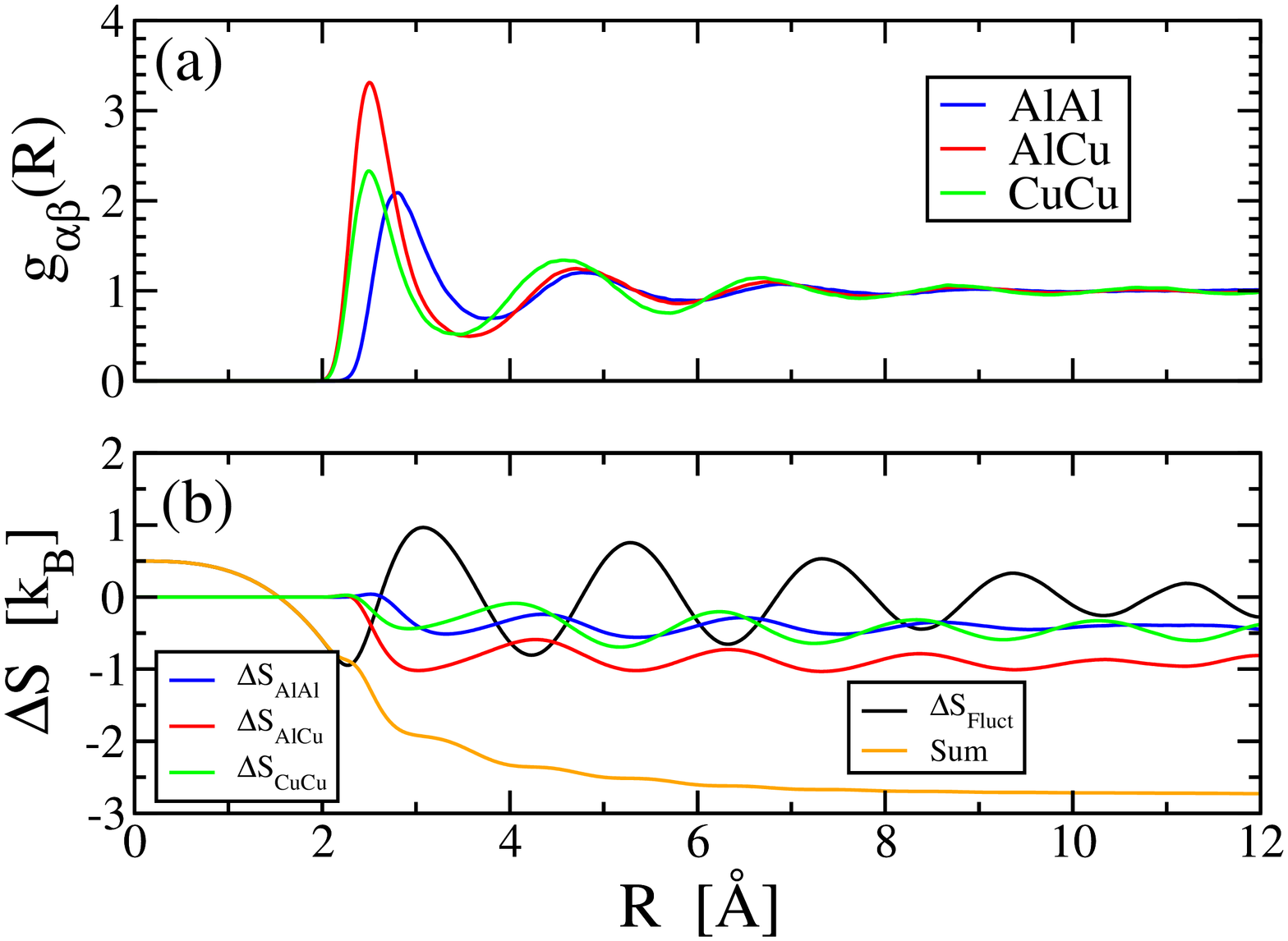}
  \caption{(a) Partial radial distribution functions $\gab(r)$ of liquid AlCu alloy at T=1373K. (b) Contributions to the entropy of liquid AlCu integrated from $r=0$ up to $R$.}
\label{fig:dS_AlCu}
\end{figure}

\begin{figure}
\includegraphics[width=5in,angle=-90]{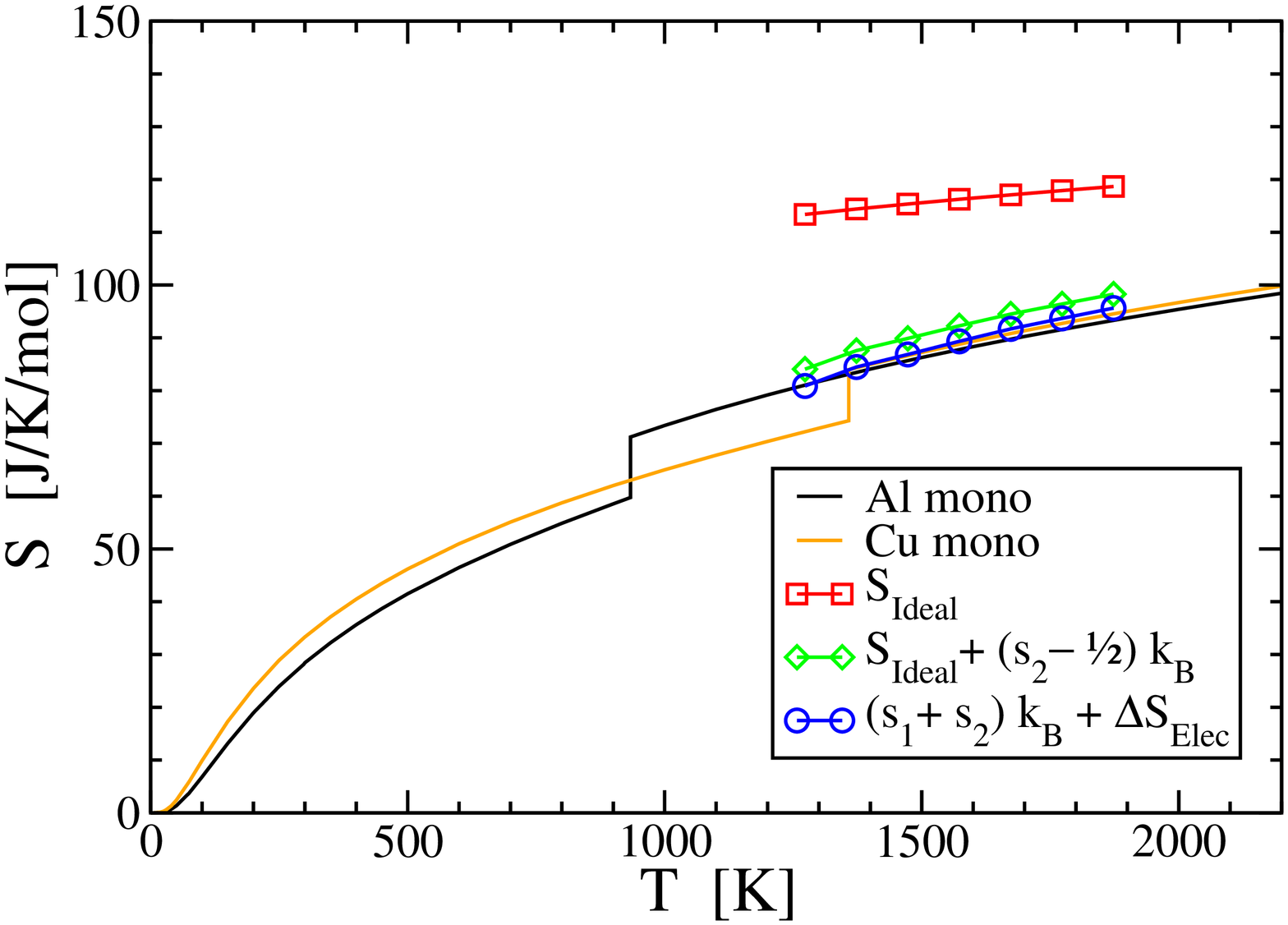}
  \caption{Entropy of liquid AlCu, comparing the experimental values of elemental Al and Cu with various approximations: the ideal gas (Eq.~(\ref{eq:S_ideal})); ideal gas with pair corrections (Eq.~(\ref{eq:S_grand})); single-body entropy with pair correction and electronic entropy, $(s_1+s_2)\kB+\Delta S_\rmElec$.}
\label{fig:S_AlCu}
\end{figure}

Simulated distribution functions and their integrals are displayed in Fig.~\ref{fig:dS_AlCu}, at T=1373K.  Note the first peak of the interspecies correlation $g_{\rm AlCu}(r)$ is much stronger than the intraspecies correlations, indicating strong chemical order.  The correlations $\gab$ reduce the entropy by $2.73~\kB$, with the interspecies Al-Cu dominating because ({\em i}) it exhibits the strongest oscillations, and ({\em ii}) it enters twice into Eqs.~(\ref{eq:S_fluct_multi}) and~(\ref{eq:S_info_multi}).  We can isolate the contribution of the average liquid structure by defining $\bar{g}(r)=\sum_\ab\xa\xb\gab(r)$ and setting
\begin{equation}
\label{eq:dSAve}
\Delta S_\rmAve\equiv\Delta S_\rmFluct[\bar{g}]
+\Delta S_\rmInfo[\bar{g}],
\end{equation}
which converges quickly to $\Delta S_\rmAve=-2.29~\kB$. Meanwhile, the contribution due to chemical order is obtained by integrating the information content contained in the relative frequencies of $\ab$ pairs~\cite{Widom16} at every separation $r$,
\begin{equation}
\label{eq:dSChem}
\Delta S_\rmChem/\kB\equiv -\frac{1}{2}\rho \sum_\ab\xa\xb\int\rmd\br~ 
\gab(r)\ln{(\gab(r)/\bar{g}(r))}.
\end{equation}
This sum converges quickly to $\Delta S_\rmChem=-0.44~\kB$, which roughly counteracts the $\kB\ln{2}$ ideal entropy of mixing.  Notice the identity $\Delta S_\rmAve+\Delta S_\rmChem=\Delta S_\rmFluct+\Delta S_\rmInfo$.

Beyond the entropy losses $\Delta S_\rmAve$ and $\Delta S_\rmChem$, there is a small additional loss of electronic entropy associated with the chemical bonding of Al and Cu, which depresses the electronic density of states at the Fermi level. At T=1373K we find values of 4.3, 3.2, and 3.1 states/eV/atom for liquid Al, AlCu, and Cu, respectively.  This results in a negative electronic entropy of mixing of $\Delta S_e=-0.02\kB$.

The entropy of the liquid alloy (see Fig.~\ref{fig:S_AlCu}) lies rather close to the average entropies of Al and Cu individually. We do not have experimental values to compare with.

\section{Conclusions}

This study demonstrates the feasibility of absolute entropy calculation based on {\em ab-initio} simulated pair correlation functions.  Given the absolute entropy, we could use the {\em ab-initio} total energies to calculate absolute free energies. Here, we focus on the reduction of entropy from the ideal gas value by the mutual information content of the pair radial distribution function. In comparison with experimental values for pure elements, we show good agreement in the case of Al and slightly worse agreement in the case of Cu. We also applied it to the case of a liquid AlCu alloy, and found that strong chemical order counteracts the ideal mixing entropy.

Two implementations of the distribution function expansion were compared, both of them truncated at the pair level. Equation~(\ref{eq:S_grand}), adds the series $1/2+1/6+\cdots=1$ to the single particle entropy $s_1$ to reach $S_\rmIdeal$, but then must subtract $1/2$ from $\Delta S_\rmFluct$, while Eq.~(\ref{eq:s1s2s3}) keeps the $1/2$ within $\Delta S_\rmFluct$. In the case of liquid Al, the latter approach yields improved agreement, as shown in Fig.~\ref{fig:S_Al}. However, in the case of liquid Cu, the former approach is favorable at low T, while the latter is best at high T. This temperature dependence is possibly due to angular correlations created by anisotropic Cu $d$-orbitals leading to a breakdown of the Kirkwood superposition approximation at low T.

Keeping the $1/2$ within $\Delta S_\rmFluct$ results in this term nearly vanishing (it is positive definite but numerically small). The fluctuation term contained in $s_3$ would likewise be small. Thus, in the approach of Eq.~(\ref{eq:s1s2s3}), $\Delta S_\rmFluct$ serves to improve convergence of the sum of integrals in Eqs.~(\ref{eq:S_fluct}) and~(\ref{eq:I}), but ultimately the entropy is primarily determined by the mutual information.

This method has been previously applied to model systems such as hard sphere and Lennard-Jones fluids, and to the one component plasma~\cite{Nettleton1958,Raveche1971b,Evans1989,Laird1992a}, as well as to simulations of real fluids using embedded atom potentials~\cite{Hoyt2000,Mendelev2014}. It can also be applied with experimentally determined correlation functions~\cite{Raveche1971b,Wallace1987,Yokoyama2002,Kelton2017}.  An analagous expansion exists for the lattice gas~\cite{Prestipino99}. Dzugutov~\cite{Dzugutov1996} utilized the method in a study reporting an empirical scaling relation between excess entropy and diffusion coefficients. However it has only rarely been applied in conjunction with {\em ab-initio} molecular dynamics~\cite{Jakse2013}. It is clear from the example of liquid Cu, as well as from the work of others~\cite{Evans1990,Laird1992a,Yokoyama2002}, that many-body terms are required to achieve high accuracy in some cases. Fortunately these are available, in principle, from AIMD.

Beyond assessing the impact of many-body terms, certain other details remain to be optimized in our calculations. We apply the PBEsol generalized gradient approximation for the exchange correlation functional because it predicts good atomic volumes (tested for solids~\cite{Perdew08,PBEsol}), but we have not tested the sensitivity of our results to other choices of functional. We consistently use systems of 200 atoms, but we have not tested the convergence of the entropy with respect to the number of atoms. After further testing and optimization, our methods could be used to develop a database of calculated liquid state entropies, both for pure metals and for alloys of interest.

Application to fluids in external fields and at interfaces~\cite{Ewing1972,Rowlinson1982,Mendelev2014} is possible by generalizing the series in Eqs.~(\ref{eq:s1_canonical}-\ref{eq:s3_canonical}) to allow for spatially varying local density $\rho(\br)$. In this case we must set
\begin{equation}
\label{eq:s1_varying}
s_1 = \frac{3}{2}-\frac{1}{N}\int\rmd\br~ \rho(\br)\ln{(\rho(\br)\Lambda^3)}.
\end{equation}
Similarly, the full, spatially varying and anisotropic, two-body density $\rho^{(2)}(\br_1,\br_2)$ (see Eq.~\ref{eq:MI}) is required in place of the translation-invariant radial distribution function $g(r)$.

\begin{acknowledgement}

MCG was supported by NETL's Research and Innovation Center's Innovative Process Technologies (IPT) Field Work Proposal and under the RES contract DE-FE0004000. MW was supported by the Department of Energy under grant DE-SC0014506. We thank R. B. Griffiths, V. Siddhu, M. Deserno and C. J. Langmead for discussions on mutual information, and R. H. Swendsen and B. Widom for discussions on liquid and gas state statistical mechanics.\footnote{Disclaimer: This project was funded by the Department of Energy, National Energy Technology Laboratory, an agency of the United States Government, through a support contract with AECOM. Neither the United States Government nor any agency thereof, nor any of their employees, nor AECOM, nor any of their employees, makes any warranty, expressed or implied, or assumes any legal liability or responsibility for the accuracy, completeness, or usefulness of any information, apparatus, product, or process disclosed, or represents that its use would not infringe privately owned rights. Reference herein to any specific commercial product, process, or service by trade name, trademark, manufacturer, or otherwise, does not necessarily constitute or imply its endorsement, recommendation, or favoring by the United States Government or any agency thereof. The views and opinions of authors expressed herein do not necessarily state or reflect those of the United States Government or any agency thereof.}

\end{acknowledgement}

\bibliography{refs}

\begin{figure}
\includegraphics[width=3in,angle=-90]{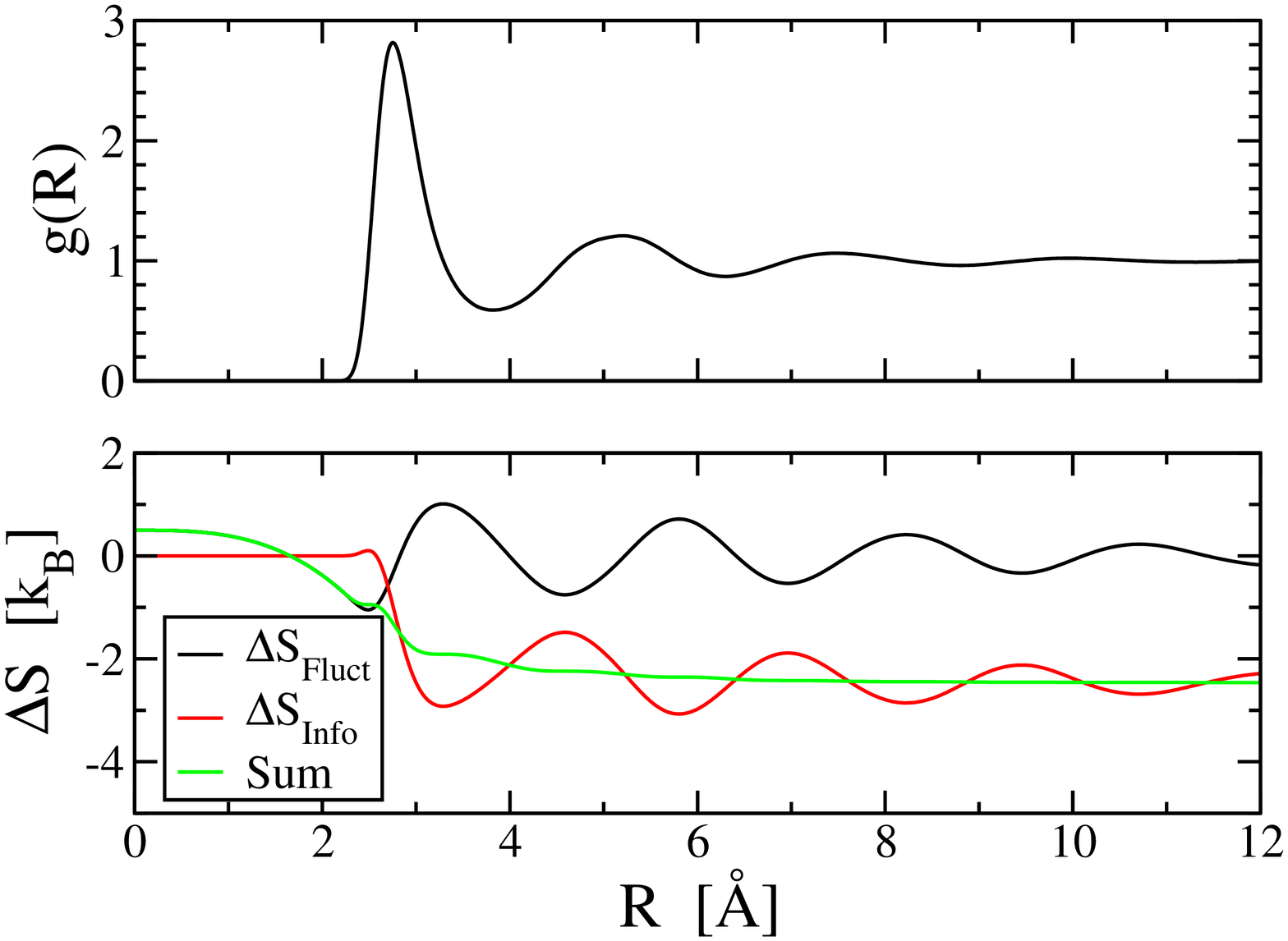}
\caption{TOC Graphic}
\end{figure}

\end{document}